# Scalable microcavity-coupled emitters in hexagonal boron nitride

Nicholas V. Proscia,[1,2] Harishankar Jayakumar,[1] Xiaochen Ge,[3] Gabriel Lopez-Morales,[1,2] Zav Shotan,[1] Weidong Zhou,[3] Carlos A. Meriles,[1,2,1] Vinod M. Menon[1,2,†]

[1]*Dept. of Physics, CUNY-City College of New York, New York, NY 10031, USA.*
[2]*CUNY-Graduate Center, New York, NY 10016, USA.*
[3]*Department of Electrical Engineering, University of Texas at Arlington, Arlington, TX 76019, USA.*

**Scalable integration of bright emitters in quantum photonic structures is an important step in the broader quest to generate and manipulate single photons via compact solid-state devices. Unfortunately, implementations relying on material platforms that also serve as the emitter host often suffer from a tradeoff between the desired emitter properties and the photonic system practicality and performance. Here, we demonstrate "pick and place" integration of a $Si_3N_4$ microdisk optical resonator with a bright emitter host in the form of ~20-nm-thick hexagonal boron nitride (hBN). The film folds around the microdisk maximizing contact to ultimately form a composite $hBN/Si_3N_4$ structure. The local strain that develops in the hBN film at the resonator circumference deterministically activates a low density of SPEs within the whispering gallery mode volume of the microdisk. These conditions allow us to demonstrate cavity-mediated out-coupling and Purcell enhancement of emission from hBN color centers through the microdisk cavity modes. Our results pave the route toward the development of scalable quantum photonic circuits with independent emitter/resonator optimization for active and passive functionalities.**

Cavity-coupled solid-state quantum emitters serve as a test bed for numerous cavity quantum electrodynamics (CQED) experiments[1-4], and are primarily utilized as a source of non-classical light[5,6] for photonic quantum information processing[7] and communication[8]. Recent practical demonstrations have exploited a variety of solid-state quantum emitter systems including quantum dots[9], single molecules[10-12], atomically thin transition metal dichalcogenides (TMDs)[13-18], and point defects in wide bandgap materials such as diamond[19-21] and silicon carbide[22]. While differing applications impose specific requirements, bright coherent sources operating at room temperature and tunable over a broad spectral range are invariably desirable. Recently, point defects in hexagonal boron nitride (hBN) have emerged as a versatile source of quantum light with unique capabilities for integration in scalable nanophotonic structures operating under ambient conditions[23]. Emitters in monolayer and multilayer hBN have been shown to emit photons at very high rates in the visible wavelength range[24,25] featuring narrow-linewidth zero-phonon lines (ZPLs)[26] and large Debye-Waller factors[23]. Current research efforts are focused on determining the defect composition[27], tunability[28], deterministic creation and emitter activation[29], and integration with photonic structures. In particular, coupling to photonic cavities, crucial for high-fidelity photon manipulation, remains an outstanding challenge.

One possible approach to cavity-coupled SPEs in hBN is via fabrication of photonic cavity structures from a bulk crystal. This technique has been successfully implemented with epitaxial quantum dots[1,2], and with color centers in diamond[4,19] and silicon carbide[22]. Recently, Kim et al.[30] extended this strategy to hBN reporting the fabrication of a photonic crystal cavity from a bulk boron nitride substrate. The two-dimensional (2D) nature of this van der Waals material, however, makes it compatible with 'pick-and-place' techniques designed to bring a thin, emitter-hosting layer in close contact with the target photonic

---

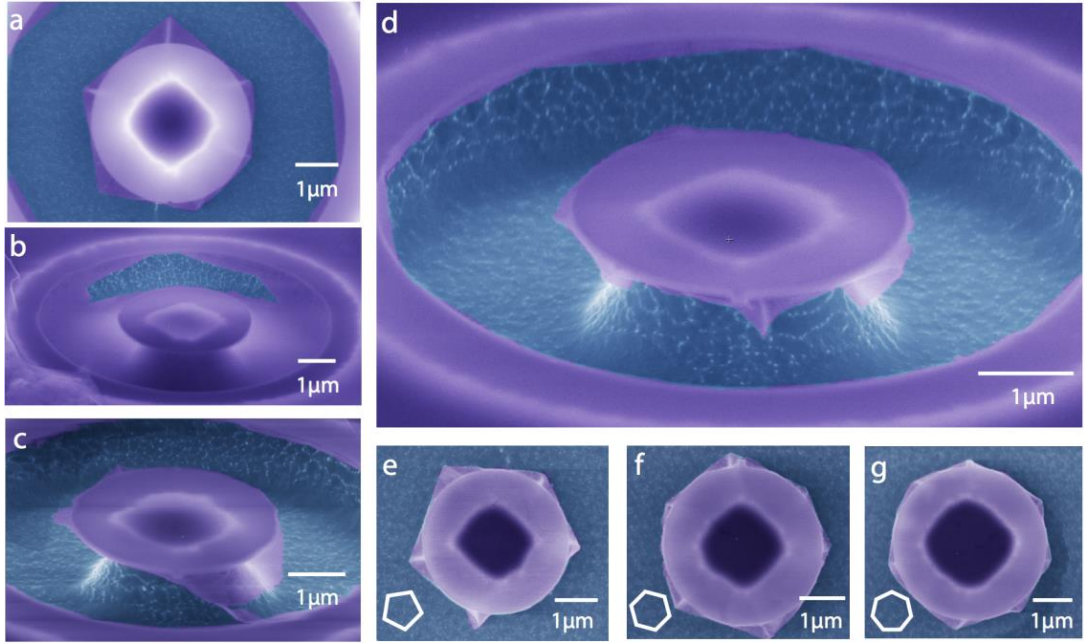

**Figure 1 | SEM imaging of 20-nm-thick hBN on Si$_3$N$_4$ microdisks. (a)** 'Tablecloth' configuration with the hBN film falling vertically on the resonator perimeter to make contact with the undercut region. **(b)** Partially torn hBN film suspended above the undercut. **(c)** Severely torn, 'tablecloth' configuration. **(d)** When separated from the rest, the hBN film wraps around the microdisk edge. **(e-g)** Top view of hBN/Si$_3$N$_4$ structures similar to that in (d) but with variable film folding geometries displaying 5, 6, and 7 edges respectively. The microdisk diameter is 3.5 μm in (a-f) and 4 μm in (g); all SEM images are false-colored for clarity.

structure. The advantage in this 'hybrid' approach is that the emitter properties (e.g., brightness, spin and/or optical lifetimes) and the photonic system performance (e.g., photon losses/quality factors) can be optimized independently. Further, the ability to choose the substrate material from a broader set circumvents complications arising from systems that cannot be produced in wafer size, are hard to process, or require fabrication steps affecting the emitter performance. These properties were recently exploited to demonstrate coupling between hBN emitters and a plasmonic resonator[31].

Here, we use a wet transfer technique to "pick and place" a ~20-nm-thick film of hBN on a silicon nitride (Si$_3$N$_4$) microdisk optical cavity. The hBN film wraps around the resonator and microdisk undercut to reproduce the local topography. Carrier capture via strain-induced potentials selectively turns the intrinsic hBN point defects at the resonator perimeter into bright emitters without the need for ion implantation or high-temperature annealing. This 'defect activation' mechanism[29] seamlessly produces emitters within the evanescent field of the microdisk cavity modes. Exploiting this singular configuration, we demonstrate cavity-mode-mediated detection of hBN color centers, and emerging coupling of the emitter ensemble to the WGM modes of the microdisk cavity.

**Results**
**Hybrid photonic microcavity.** We use a wet transfer technique[29] to overlay a ~20-nm-thick flake of hBN on an array of whispering gallery mode (WGM) resonators in the form of silicon nitride (Si$_3$N$_4$) microdisks, (Fig. 1). These cavities are designed to sustain in-plane transverse electric (TE) WGMs in the visible wavelength range between 550 to 650 nm, roughly matched to the dipole orientation[32] and emission wavelength of the hBN emitters. All microdisks feature a tip on its periphery to out-couple the scattered resonator modes into a high numerical aperture (NA) confocal microscope objective (Supplementary Figure 1). The cavity mode $Q$-factors are found to be of order ~1000 (~3500) for the 3 μm (4 μm) diameter resonators; similarly, the cavity free spectral range (FSR) goes from ~20 nm to ~15 nm within the investigated spectral band. Further details on the microdisk fabrication and optical properties can be found in Supplementary Note 1.

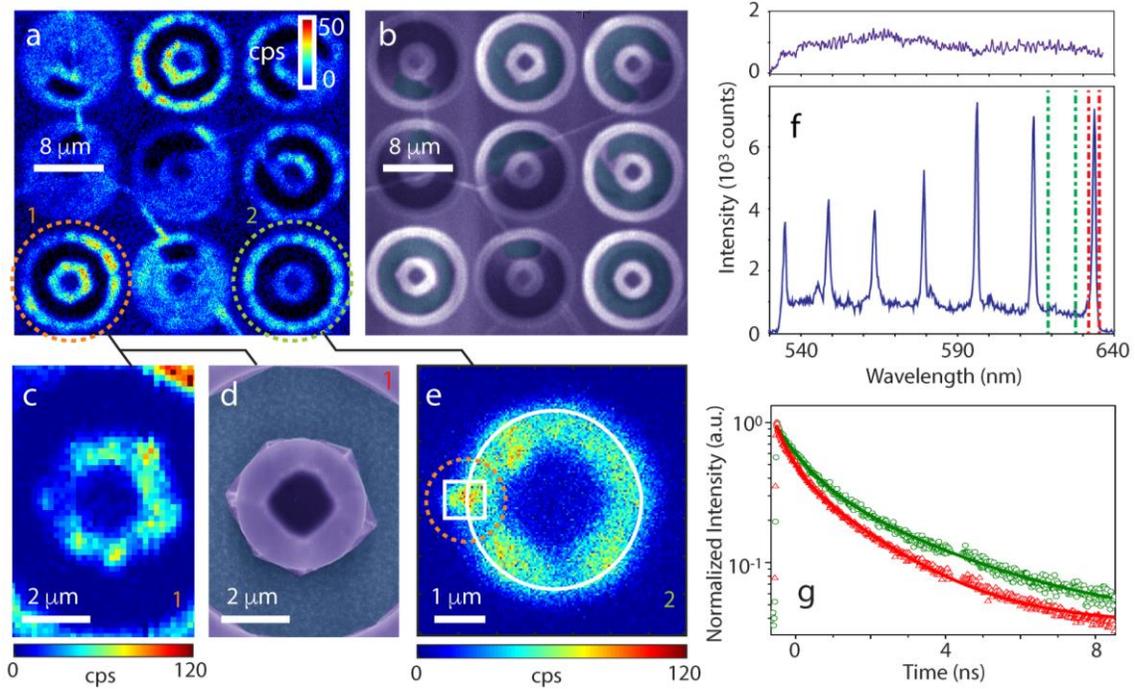

**Figure 2 | Photoluminescence from strain activated emitters in hBN. (a)** Confocal fluorescence image of an array of hBN/Si$_3$N$_4$, 3.5-μm-diameter microdisk cavities upon 510 nm laser excitation. At 'wrapped' resonator sites, emission preferentially stems from the microdisk perimeter and the outer contour of the undercut etch. **(b)** SEM image of the array shown in (a). **(d, c)** Zoomed confocal and SEM images of microdisk resonator 1 (lower left corner in (a)). **(e)** Same as in (d) but for resonator 2 (right lower corner in (a)). Overlaid white circle (square) serves as a guide for the microdisk perimeter (in/out-coupler location). **(f)** Photo-luminescence spectra at the in/out-coupler or from another point at the disk periphery upon 520 nm, 1 mW laser excitation (upper and lower plots, respectively). **(g)** Excited state lifetime measurements in emitters resonant with or detuned from the cavity modes (respectively, spectral windows between dashed red or dashed green vertical lines in (f)). Solid red and green traces indicate three-level exponential fits in either case, respectively.

Upon hBN transfer to the resonator array, the composite hBN/Si$_3$N$_4$ structures that emerge take different geometries, as illustrated in the scanning electron microscopy (SEM) images of Figure 1. For example, on occasions we find that the flake falls almost vertically at the microdisk perimeter to make contact with the undercut etch without noticeable damage ('tablecloth' geometry in Figure 1a), while at other times the film bridges the space between the microdisk and the surrounding resonator host crystal without undergoing any visible distortion ('suspended' geometry in Figure 1b). Typically, the unsupported section of the film undergoes variable rupture at the resonator periphery (see, for example, Figures 1b and 1c), ultimately tearing apart completely to fold around the microdisk ('wrapped' geometry in Figure 1d). This type of hBN wrapping — seen in approximately one third of the microdisks — features a polygonal shape with 5 to 7 vertices, the number of corners typically growing with the disk diameter (Figures 1e to 1g).

Out of all configurations, here we focus on the latter, 'wrapped' type where emitters tend to cluster near the disc periphery and thus lie proximal to the evanescent field of the WGM modes. This is shown in the scanning fluorescence image of Figure 2a, predominantly brighter in areas around the resonator circumference and outer contour of the undercut. Negligible photo-luminescence (PL) is seen in sections where the hBN film is torn, hence confirming photon emission originates exclusively from the guest, not the substrate material (Supplementary Figure S2). On the other hand, suspended films feature comparable levels of background fluorescence both in the supported and unsupported sections, thus indicating the substrate does not have a major effect on the emitter properties[33].

We recently observed similar phenomenology in hBN films on arrays of $SiO_2$ micro-pillars — where bright color centers cluster exclusively at the pillar sites — and showed this response can be understood as a form of emitter activation driven by carrier capture in strain-induced potentials[29]. This same process is at work here, selectively activating the emission from color centers near the resonator's periphery (particularly the vertices of the polygonal folds) where the hBN curvature and thus the accumulated strain is highest (see zoomed confocal image of Figure 2c and isometric SEM image in Figure 2d). We contrast this type of emitter engineering with that attained via hBN annealing at ~850 $^0$C (Supplementary Figure 3), broadly used but largely non-selective in its impact on emitters, and generally ill-adapted to temperature-sensitive structures.

**Interplay between cavity modes and color center emission.** The impact of the WGMs on the emitters' dynamics is not apparent when inspecting the fluorescence brightness, often times uniform along the disk perimeter (Figure 2e). It can be clearly exposed, however, as we compare the spectral and temporal properties of light collected at or away from the photon out-coupler at the disk circumference. The spectra in Figure 2f provide one first demonstration, here in the form of evenly-spaced, narrow lines exclusively present in the light emission from the out-coupler site. Taking into account the broad, heterogeneous spectra typical of emitter ensembles in strained hBN[29,34], we associate this pattern to contributions from color centers within the ensemble whose zero-phonon-line (ZPL) is resonant with one of the resonator modes. Naturally, emitters of this class populate uniformly the disk perimeter, but their emission becomes selectively observable at the out-coupler, purposely designed to direct light trapped within the resonator into the objective.

Accompanying the WGM pattern we find a non-negligible 'background' component, which we attribute to isotropic color center emission not mediated by the resonator. This contribution is analogous to that found on the disk perimeter away from the scatterer (upper plot in Figure 2f), or in other activated regions of the film removed from the resonators (e.g., outer undercut edge, not shown). We confirm the distinct nature of these two contributions by examining the fluorescence temporal decay after pulsed excitation; to separately gauge one component or the other we restrict photon collection to spectral bands overlapping or not with a resonator mode (respectively, dashed red or dashed green spectral windows, lower half of Figure 2f). The excited state lifetime of resonant color centers (which we characterize via the smaller time constant in a multi-exponential fit) is approximately 30% shorter than that of detuned emitters, indicative of incipient coupling to the microdisk cavity (Figure 2g). Intriguingly, this lifetime-derived Purcell enhancement $P_{meas} = 1.3$ is significantly lower than the anticipated value $P_{calc} = 8.4$, as calculated from the ~1 nm full width at half maximum of the measured cavity modes (cavity $Q$-factor of ~650). The reasons are presently unclear, but we hypothesize that it may partly stem from imperfect replication of the microdisk topography (hence leading to a lower field mode overlap with the emitter) as well as emitter misalignment with the TE cavity mode.

Extending the observations above, the experiments in Figure 3 alter our confocal imaging scheme — so far limited to collecting fluorescence from the point of illumination — to one where the excitation and observation loci can be adjusted freely (Figure 3a). As a first demonstration, we set the collection point at the out-coupler and scan the laser beam across the $Si_3N_4$/hBN structure hence allowing us to collect the light scattered from the resonator tip as the laser excites the emitters at other points of the microdisk. Figure 3b reproduces the image from a 4-μm-diameter resonator emerging from this non-local excitation scheme: Away from the out-coupler area (orange square), we identify several bright spots across the disk periphery, likely connected to points in the resonator where the hBN film wrinkles or folds, hence better allowing the excitation beam to couple into the cavity. The corresponding spectra prominently feature the WGM wavelengths with reduced off-resonance contributions (insert Figures (*i*) through (*iv*) in 3b). These observations contrast with those obtained at the same sites via standard confocal microscopy (i.e., co-local excitation and collection), characterized by broad, non-resonant photo-luminescence (right inserts in Figure 3b). In Figure 3c we use a variant of the above technique where the excitation and collection focal points move jointly, with a fixed separation matched to the disk diameter; upon properly adapting the scanning trajectory, light collected at the scatterer exclusively stems from excitation at the opposite side of the disk. The resulting disk-mediated spectrum shows virtually no background fluorescence (right insert in Figure 3c).

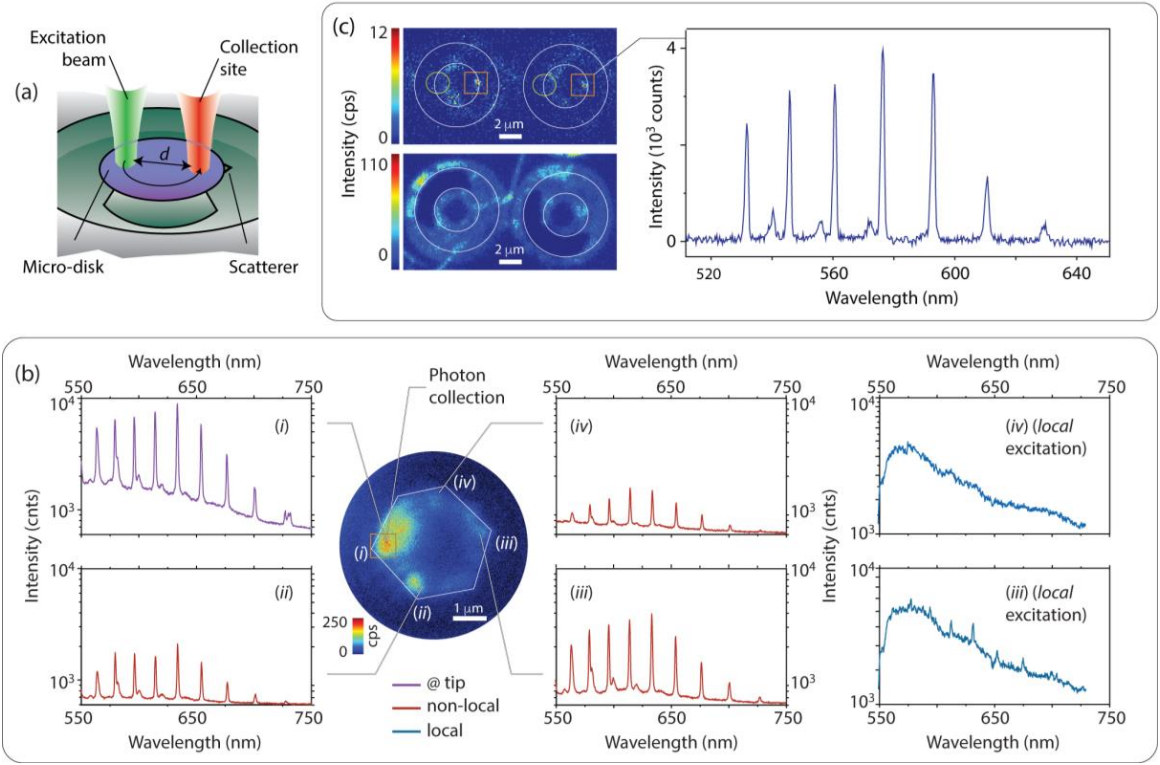

**Figure 3 | Resonator-mediated photon emission from color centers in hBN. (a)** Non-local excitation schematics. **(b)** (Main) Fluorescence image of an hBN-wrapped micro-disk using the scheme in (a); in this case, photons are collected exclusively at the tip site (orange square, point (*i*) in the image) while the laser beam scans the rest of the sample. (*i*) Fluorescence spectrum at site (*i*), where photon excitation and collection coincide. (*ii-iv*) Fluorescence spectra upon optical excitation at sites (*ii*), (*iii*), and (*iv*), respectively. Background contributions are considerably suppressed. (Right) Fluorescence spectra at sites (*iii*) and (*iv*) obtained via standard confocal microscopy (i.e., co-local excitation and collection). **(c)** (Top) Same as in (b) but for the case where the collection point moves jointly with the excitation point, at a distance $d$ nearly matched to the micro-disk diameter. Photons detected at the resonator scatterer (red square) originate from emitters in the opposite side of the disk (green oval). White traces demark the contours of the resonators and surrounding undercut edges. (Bottom) Standard confocal image of the same area. (Right) Fluorescence spectrum at the out-coupler site featuring near-complete background suppression.

While the experiments above rely on heterogeneously broadened ensembles, we occasionally observe unusually bright, spatially localized fluorescence suggesting the presence of individual emitters. Likely associated to a different class of point defects, these 'super-emitters' feature relatively narrow ZPLs (~5-10 nm wide), and thus stand out over the 'background', featureless fluorescence of the ensemble. Unfortunately, photon correlation measurement proved impossible to implement, not only because these emitters are rare and tend to bleach after comparatively short time intervals, but also because the ensemble provides a large background signal we cannot suppress. On the other hand, these color centers are prone to strong spectral diffusion (possibly due to photo-induced charge fluctuations during illumination), which gives us the opportunity to monitor their response as they transiently become resonant with one of the WGMs. One example is shown in Figure 4a, where we display a time series of photo-luminescence spectra acquired every second during a 30-s-long window. In this first case, the super-emitter ZPL is seen to hover near the cavity mode at 548 nm, slowly moving from higher to lower wavelengths; we don't find, however, major changes in the mode amplitude, hence suggesting weak coupling to the cavity. By contrast, the observations in Figure 4b reveals a super-emitter undergoing fast spectral diffusion over the entire spectral window; emitter blinking (seen, e.g., near half the probed time span), leads to simultaneous darkening of all cavity modes, suggesting this color center plays a dominant role in the fluorescence from this micro-disk. The disproportionate enhancement of the optical resonances at 568 nm and 583 nm (accompanied by line broadening, see also Figure 4c) hints at efficient single-color-center coupling to the cavity. In the absence of photon correlation measurements, however, these observations must be treated cautiously.

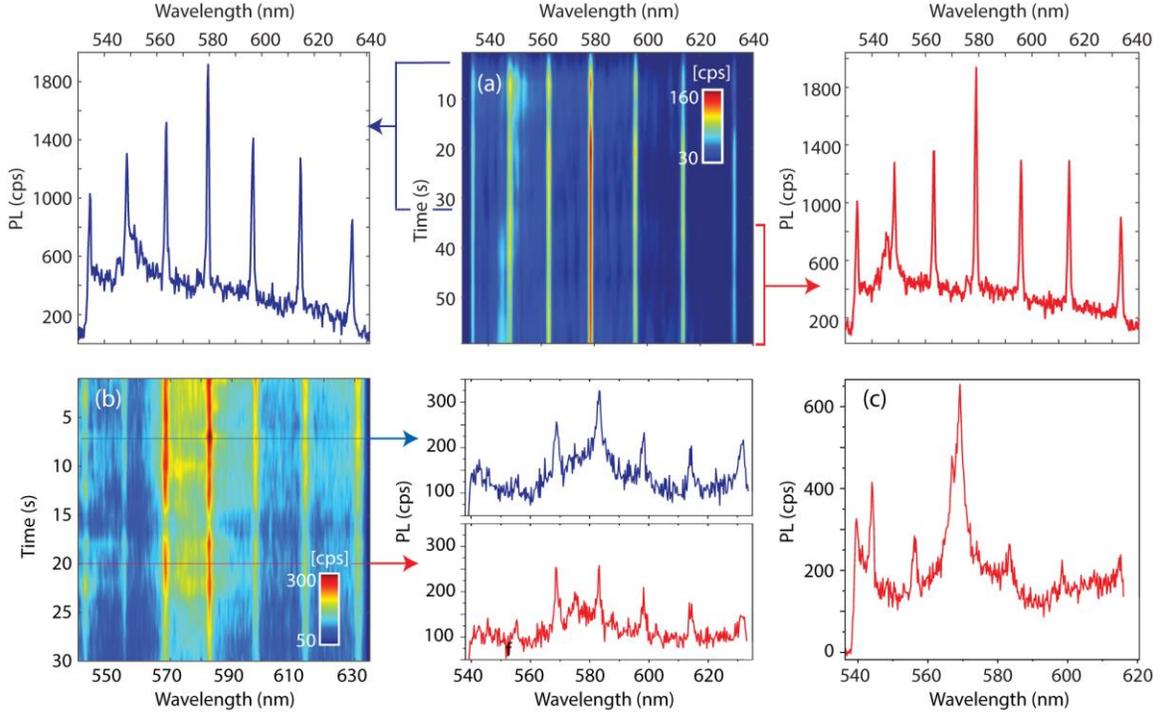

**Figure 4 | Transient spectroscopy of hBN 'super-emitters' and microdisk mode coupling. (a)** (Center) Set of fluorescence spectra from a microdisk with a super-emitter undergoing spectral diffusion; the acquisition time per spectrum is 1 s. The left (right) plot shows the time-integrated spectrum from 6 to 32 s (38 to 58 s). **(b)** Same as in (a) but for another micro-disk where the super-emitter is dominant. The blue (red) spectrum to the right of the plot is a cross section of the main plot at 7 s (20 s). Notice the emitter blinking at times near 17 s. **(c)** Spectrum of the same super-emitter at a later time, overlapping with the cavity mode at 568 nm (not shown in the time trace). The measurement conditions are those of Figure 2f.

**Discussion**

In summary, our work demonstrates a new, alternative route to controlling the emission of cavity-integrated color centers that capitalizes on the unique versatility of two-dimensional materials. In the composite structures we produce, the 2D system and optical cavity fuse to form a single unit, where color centers self-activate in areas coincident with those occupied by the resonator modes. The latter allows us to show cavity-mediated photon emission and emergent color-center ensemble coupling to the cavity. These results represent, therefore, a first step in the quest to developing practical hybrid, chip-scale quantum photonic systems, where the properties of the optically active point defect, their 2D material host, and the supporting, photon-guiding structure can be independently optimized. Indeed, we anticipate several possible extensions, for example, through the use of more compact, smaller-mode-volume beam resonators as a strategy to enhance emitter/cavity coupling, and attain good multiphoton suppression and anti-bunching response. By the same token, combining strain-activation and defect engineering (e.g., via focused ion implantation or electron bombardment) may provide a means to generate optically active, spectrally stable emitters at precise locations in otherwise pristine 2D hosts. Finally, extensions to structures comprising tapered fibers or on-chip waveguides evanescently coupled to the resonators should serve as a platform for high-photon extraction and a variety of cavity quantum electrodynamics experiments.

## Methods

**Microdisk Fabrication.** The microdisk (MD) was fabricated from a 112 nm $Si_3N_4$ layer on a Silicon substrate. The stoichiometric silicon nitride film was deposited on the silicon substrate by low pressure chemical vapor deposition (LPCVD). The MD structures were patterned on ZEP520A resist by electron beam lithography (EBL) and transferred onto the $Si_3N_4$ layer by inductively coupled plasma reactive ion etching (ICP-RIE) using a gas combination of SF6, CHF3 and He. The silicon substrate was then undercut by another ICP-RIE process with SF6 to form the suspended $Si_3N_4$ microdisk structure. Finally, the remaining e-beam resist was removed by N-Methyl-2-pyrrolidone (NMP) followed by $O_2$ plasma ashing.

**Experimental setup.** The optical characterization of the sample was performed with a custom-built confocal microscope in the collinear excitation collection geometry. A piezo Nano stage was used to raster scan the sample across a fixed laser excitation and collection path. An infinity-corrected 50x, 0.83 NA Olympus objective was used to collect the micro-PL. The excitation source was a 500 fs pulsed fiber laser with a repetition rate of 80 MHz operating at 510 nm (Toptica FemtoFiber pro TVIS); the focused spot size had a diameter of ~1 μm.  A 532 nm angle-tuned laserline filter (Thorlabs FL532-10) was used to filter unwanted spectral modes of the laser. Angle-tuned (edged from ~540 to ~640 nm) shortpass and longpass filters (Semrock TSP01 & Semrock TLP01, respectively) were used to cut off the reflected laser excitation as well as selectively isolate the MD cavity modes for further analysis.  Lifetime measurements were performed using an avalanche photo diode (APD) (Micro Photon Devices PDM series) with 50 ps jitter, and a Picoquant – Picoharp 300 time tagger.  An 80-20 emission beam-splitter provided real-time spectral analysis with an iHR-320 Horiba spectrometer.

## Author Contributions
N.V.P., H.J., G.L-M., and Z.S., carried out the experiments with the supervision of C.A.M. and V.M.M. The microdisk resonators were designed and fabricated by X.G. under the direction of W.Z. All authors discussed and wrote the manuscript.

## Acknowledgements
We thank Dr. Daniela Pagliero and Dr. Jacob Henshaw for technical assistance with part of the instrumentation used. N.V.P., G.L.M., and V.M.M. acknowledge support from the NSF MRSEC program (DMR-1420634) and the NSF EFRI 2-DARE program (EFMA -1542863). H.J., G.L.M., and C.A.M acknowledge support from the National Science Foundation through grants NSF-1619896, NSF-1726573, and from Research Corporation for Science Advancement through a FRED Award. All authors acknowledge support from and access to the infrastructure provided by the NSF CREST IDEALS (NSF-1547830) and the CUNY-ASRC Nanofabrication Facility.

Supplementary Information for

# Scalable microcavity-coupled emitters in hexagonal Boron Nitride

Nicholas V. Proscia,[1,2] Harishankar Jayakumar,[1,] Xiaochen Ge,[3] Gabriel Lopez-Morales,[1,2] Zav Shotan,[1] Weidong Zhou,[3] Carlos A. Meriles,[1,2,†] Vinod M. Menon[1,2,†]

[1]*Dept. of Physics, CUNY-City College of New York, New York, NY 10031, USA.*
[2]*CUNY-Graduate Center, New York, NY 10016, USA.*
[3]*Department of Electrical Engineering, University of Texas at Arlington, Arlington, TX 76019, USA.*

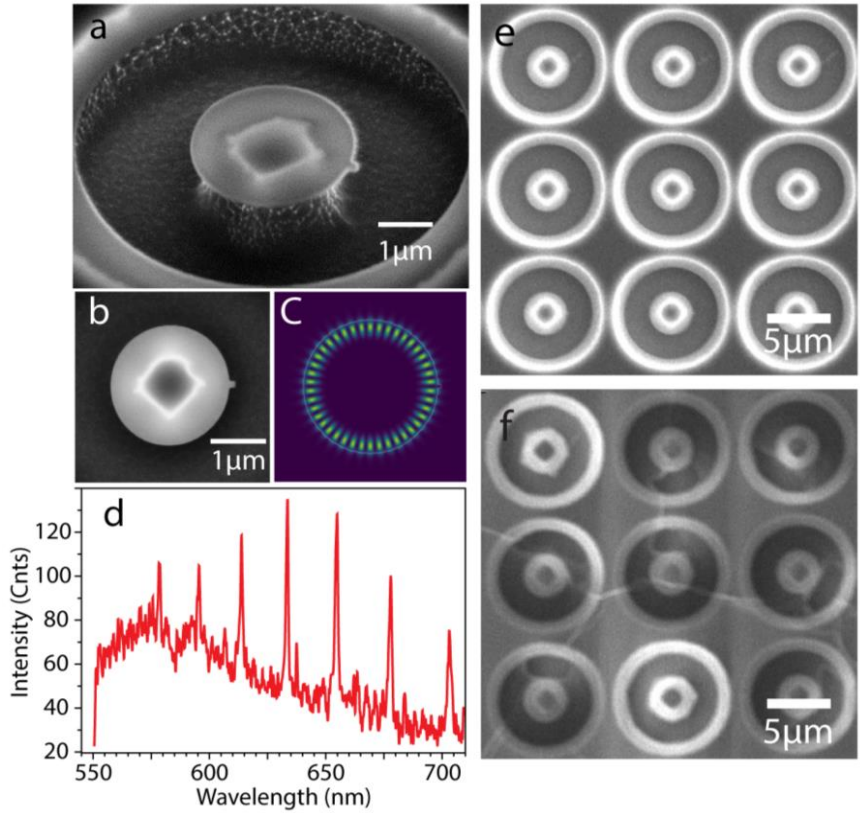

**Supplementary Figure 1 | Bare microdisk characterization** (a) Angled SEM image of 1.5-µm-radius $Si_3N_4$ microdisk with a tip to out-couple the cavity modes. (b) A top view SEM image of the microdisk in (a). (c) The field plot at 632 nm of a 1.7-µm-radius tipped microdisk in (a). (d) Fluorescence spectrum from a bare 1.7-µm-radius $Si_3N_4$ microdisk. The cavity modes become observable after an integration time of 180 secs. By comparing to the spectra in the main text (obtained after 1 sec integration), we conclude the intrinsic microdisk fluorescence is negligible. (e, f) SEM image of the micro-disks before and after hBN transfer, respectively.

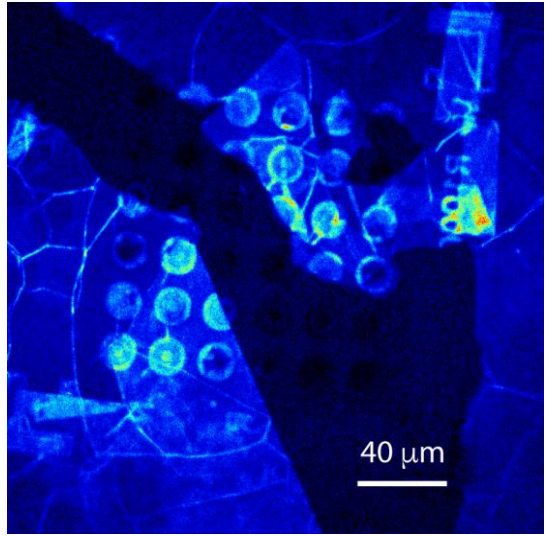

**Supplementary Figure 2 | PL scan of a microdisk array and overlaid hBN**. Dark areas correspond to sections of the array lacking the hBN film. The high contrast between the bright and dark sections of the image shows that the contribution from the bare micro-disks to the observed photoluminescence is negligible.

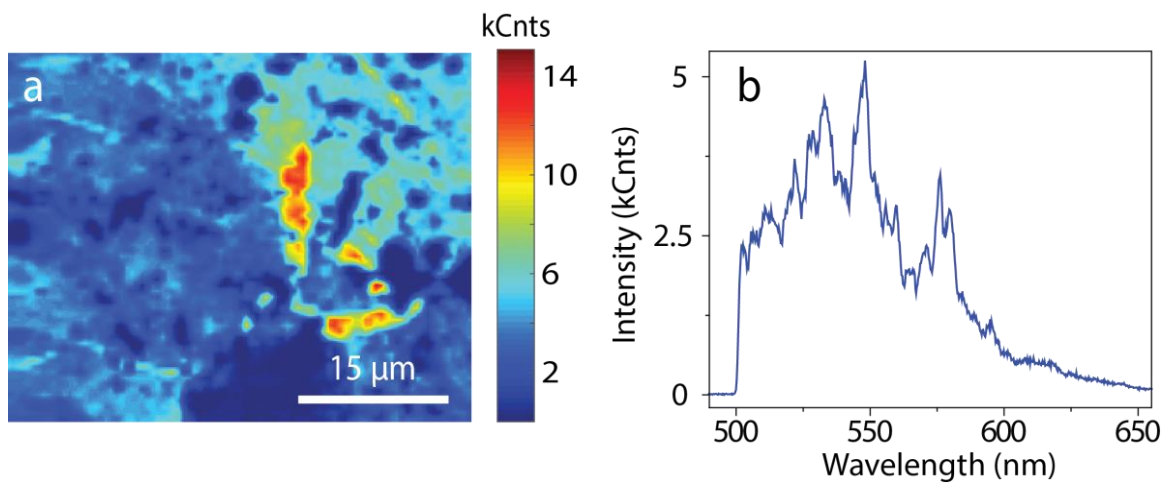

**Supplementary Figure 3 | Defect emission from annealed sample.** (a) Photoluminescence image of a CVD-grown hBN sample similar to those used in the main text after a 30-minute anneal at 850 $^0$C. A significant percentage of the h-BN film photo-luminesces. (b) A representative spectrum of the emission found in (a).

**Supplementary Note 1: Characterization of Bare Microdisk cavities**

The $Si_3N_4$ microdisk (MD) resonators were fabricated with radii ranging from 1500nm-2000nm. The MDs were designed for the electric field TE modes to be parallel with in-plane dipoles associated with the hBN film[1,2]. In addition, for each MD radius, corresponding disks were constructed either without an out-coupling structure or with a notch or tip on the periphery to help scatter out the resonator modes for collection via perpendicular microscopy (Supplementary Figures 1a and 1b). It is found that the plain and notched MDs were extremely poor for out-coupling the modes into the objective and thus only the tipped MDs were investigated.

The *Q*-factor of a bare tipped disk with radius 1500 nm was measured to be ~2000 and varied depending on the radius of the disk reaching up to ~3500 for the largest disks (2000 nm radius). The free spectral range (FSR) of the disks ranged from ~20 nm for the 1500-nm-radius disks to ~15 nm for the 2000-nm-radius disks within the investigated spectral band (540 nm-640 nm). Spectra of the tipped 1500-nm- and 2000-nm-radius MDs is shown in Supplementary Figure 1d. The bare MD modes were probed via weak PL generated from the $Si_3N_4$, about two orders of magnitude weaker than the hBN emission and hence not detectable on the measurement time scales used to study the hBN emission (Supplementary Figure 2).

**Supplementary Note 2: Defect activation and optical characteristics**

In our samples, point defects are formed during growth and require activation either via strain or through annealing. Supplementary Figure 3a shows a confocal photo-luminescence (PL) image of a 20-nm-layer hBN on a $SiO_2$ substrate after annealing for 30 mins at 850 $^0$C in an argon environment at 1 Torr. We find that most of the hBN film exhibits bright PL and the emission spectrum (Supplementary Figure 3b) is comparable to the emission produced at strained locations.

The time dynamics of the defect emission can vary greatly from emitter to emitter. Most emitters were found to go from a bright to dark state (i.e., to blink) intermittently on a timescale ranging from milliseconds to seconds. In addition, it is found that the ZPLs show spectral diffusion from ~10 to ~100 nm over the course of a measurement, which can be seen in Figure 4 in the main text. We find that a fs pulsed laser operating at 510 nm increases the stability of the emitters.

**Supplementary References**